\documentclass[10pt,journal]{IEEEtran}
\usepackage{cite}
\usepackage{amsmath,amsfonts}
\usepackage{amssymb}
\usepackage{graphicx}
\usepackage{color}
\usepackage{subcaption}
\usepackage{multirow}
\usepackage[ruled, vlined, linesnumbered]{algorithm2e}
\usepackage{caption}
\usepackage{booktabs}
\begin{document}

\title{A Secure and Efficient Distributed Semantic Communication System for Heterogeneous Internet of Things}

\author{
	
	Weihao Zeng,~\IEEEmembership{Graduate Student Member,~IEEE,}
	Xinyu Xu,
	Qianyun Zhang,~\IEEEmembership{Senior Member,~IEEE,}
	Jiting Shi\\
	Zhenyu Guan,~\IEEEmembership{Member,~IEEE,}
	Shufeng Li,~\IEEEmembership{Member,~IEEE,}
	Zhijin Qin,~\IEEEmembership{Senior Member,~IEEE}
        
\thanks{W. Zeng, X. Xu, Q. Zhang, J. Shi and Z. Guan are with the School of Cyber Science and Technology, Beihang University, Beijing 100191, China (email: \{zengweihao, xuxinyu, zhangqianyun, shijiting, guanzhenyu\}@buaa.edu.cn).}
\thanks{Shufeng Li is with the State Key Laboratory of Media Convergence and Communication, Communication University of China, Beijing 100024, China (e-mail: lishufeng@cuc.edu.cn).}
\thanks{Zhijin Qin is with the Department of Electronic Engineering, Tsinghua University, Beijing 100084, China (e-mail: qinzhijin@tsinghua.edu.cn). }

}

\maketitle

\begin{abstract}
Semantic communications are expected to improve the transmission efficiency in Internet of Things (IoT) networks.
However, the distributed nature of networks and heterogeneity of devices challenge the secure utilization of semantic communication systems.
In this paper, we develop a distributed semantic communication system that achieves the security and efficiency during update and usage phases.
A blockchain-based trust scheme for update is designed to continuously train and synchronize the system in dynamic IoT environments.
To improve the updating efficiency, we propose a flexible semantic coding method base on compressive semantic knowledge bases.
It greatly reduces the amount of data shared among devices for system update, and realizes 
the flexible adjustment of the size of knowledge bases and the number of transmitted signal symbols in model training and inference stages.
In the usage phase, a signature mechanism for lossy semantics is introduced to guarantee the integrity and authenticity of the transmitted semantics in lossy semantic communications.
We further design a noise-aware differential privacy mechanism, which introduces optimized noise based on the different channel information available to heterogeneous devices.
Experiments on text transmission tasks show that the proposed system achieves the protection of the integrity and privacy for exchanged semantics, and reduces the data to be transmitted in the update phase by about $35\%$ to $88\%$, and in the usage phase by $60\%$ compared with related works.

\end{abstract}

\begin{IEEEkeywords}
Internet of Things, semantic communications, security and privacy.
\end{IEEEkeywords}
\section{Introduction}
\IEEEPARstart{T}{he} 
proliferation of the Internet of Things (IoT) has led to a significant increase in data volumes and network connectivity. This rapid expansion highlights the necessity for efficient communication systems within IoT networks.
Semantic communications are novel communication paradigms that focus on directly conveying intended meanings and sharing only the essential information relevant to the receiver's needs, i.e. semantics\cite{qin2021semantic,gunduz2022beyond}. 
Semantic communication codecs are built on neural network models and shared knowledge bases, effectively extracting semantic features from diverse sources and accurately interpreting them to facilitate execution of specific tasks.
It has emerged as a promising approach to enhance transmission capabilities of IoT devices, and pave the way for more intelligent IoT tasks\cite{xie2020lite,du2023rethinking}. 

However, the distributed nature of IoT networks pose significant challenges to the practical deployment of semantic communication systems.
Unlike end-to-end semantic communications\cite{xie2021deep}, distributed systems within IoT networks require more complex multi-party interactions for the system update.
To be specific, there are two keys to system update: system synchronization and training.
The system synchronization means matching semantic communication codecs among multiple participants to prevent inaccurate extraction and interpretation of semantic information.
On the other hand, ever-changing communication tasks in dynamic IoT scenarios necessitate ongoing training of semantic communication codecs.
This requires IoT devices to collect evolving task-related data which is inevitably distributed across different devices.
Then, these devices perform local update training and interact for collaborative training, such as federated learning\cite{mcmahan2017communication}, to exploit the distributed data.

To establish a secure distributed semantic communication system, the first challenge is to protect the integrity and availability of exchanged data during system updates.
Its integrity is threatened by various attacks, such as data tampering, data falsification and man-in-the-middle attacks\cite{deogirikar2017security}. Adversaries can maliciously modify or falsify the information exchanged during system synchronization, mismatching models and knowledge bases among devices. 
They also impede the convergence of models and the representation of knowledge bases by introducing perturbations into the information related to the system training\cite{feng2021bafl}.  
In addition, protecting the availability of the data for system update also presents a challenge, given the inherent dynamics of IoT network topology and the potential for device malfunctions, disconnections, and communication delays and so on\cite{howard2015raft}.
External attacks, such as distributed denial-of-service attacks, also compromise the availability of the data\cite{meneghello2019iot}.
To address above security issues in the system update, it is important to develop a 
trustworthy scheme for securely updating distributed semantic communication systems.

On the basis of ensuring security, further improving the efficiency of system update is another key issue for heterogeneous devices in IoT networks.
Heterogeneous devices have diverse transmission and computation capabilities.
During the system update, the direct exchange of entire models and knowledge bases among IoT devices imposes severe burdens on these transmission-limited IoT devices.
This is due to the substantial size of the current implementation of models\cite{xie2020lite} and knowledge base, such as knowledge graph\cite{zhou2023cognitive,jiang2022reliable}, training datasets\cite{zhang2022deep} and feature vector sets\cite{sun2023semantic}, which result in overwhelming data transmission.
In addition, the immense data size of models and knowledge bases significantly increases the computational overhead during the model training and inference.
This challenge is particularly acute for IoT devices with limited computing power, leading to high communication latency and reduced system efficiency.
Therefore, it is imperative to develop a semantic coding method that facilitates efficient synchronization, training of the system, and possess flexibility to accommodate a diverse range of devices.

During the use phase, the transmitted semantics is also vulnerable to integrity and authenticity threats as system update data.
The lossy transmission nature of semantic communications exacerbates difficulties in protecting the integrity and authenticity of the semantics exchanged.
The processes of extracting and interpreting the semantics by neural networks introduces model noise, and the channel noise is added when semantic information is transmitted through wireless channels.
Traditional digital signature mechanisms cannot be directly applied to semantic communications, because any small distortions introduced into the semantic information lead to the verification failure\cite{pub2000digital}.
Therefore, the system urgently requires a signature mechanism oriented towards lossy semantics to verify that the transmitted semantics has not been tampered with or forged.

Although semantic communications deliver only the semantics and keep the raw data local, thus limiting the exposure of individual data to other parties, privacy concerns remain acute.
The sensitive information in the original data remains implicit in the semantics and can be inferred by methods such as model inversion attacks\cite{chen2023model, zhang2020secret}, and data inference attacks\cite{ye2022one}.
For ensuring privacy in data analysis and utilization, differential privacy (DP)\cite{dwork2006differential, abadi2016deep} has emerged as a prominent framework. It provides a rigorous mathematical defend against data inference attacks.
The differential privacy is achieved mainly by adding carefully designed noise to reduce the significance of the data distribution.
Considering the similarity between differential privacy implementations and semantic communication lossy processes, it is meaningful to jointly analyze differential privacy noise and semantic communication process noise, and  develop a less noise-adding differential privacy mechanism in semantic communications to achieve privacy preservation.

To tackle above challenges presented in semantic communications within IoT networks, we propose a secure and efficient distributed semantic communication system. Our contributions are presented in detail as follows.
\begin{enumerate}
	\item{
		We propose a blockchain-based trusted update scheme for the distributed semantic communication system.
		In this scheme, codecs and update-related information are shared among devices in an integrity-preserving manner.
		Furthermore, the availability of the distributed system in complex and changing IoT networks is guaranteed.
	}
	\item{
		To improve the efficiency of system update, a flexible semantic coding method based on compressive semantic knowledge bases is proposed.
		By mainly updating and synchronizing semantic knowledge bases, the scheme significantly reduces the amount of data that needs to be exchanged.
		Meanwhile, it provides heterogeneous IoT devices with the flexibility to adjust the size of the knowledge base and the number of transmitted signal symbols in model training and inference stages.
	}
	\item{
		We design a signature mechanism for lossy semantics to verify whether the received semantics has been tampered with or forged.
		The mechanism addresses the challenge of verifying the integrity and authenticity of semantic information in lossy communications by signing small samples of symbols and transmitting them in error-free links.
		The effect of wireless channel on the transmitted semantics is investigated to ensure the completeness of the signature mechanism.
	}
	
	\item{
		We introduce a noise-aware differential privacy mechanism to uniformly and transparently provide differential privacy protections in any semantic communication tasks.
		Taking into account distortions caused by wireless channels and model, the mechanism optimally adds noise into signal symbols to defend against malicious data analysis.
		To enhance the adaptability of the mechanism, various capabilities of heterogeneous devices to estimate the channel information and the model noise are analyzed in the mechanism design.
	}

\end{enumerate}

The rest of this article is organized in the following way. 
The related work is presented in Section II.
We present system model in Section III including scenario description, semantic communication system model with compressive semantic knowledge base and problem definition.
Section IV introduces an overview of the proposed system, followed by a detailed description of four important mechanism.
The performance of the system are evaluated in Section V.
Finally, we conclude our work in Section VI.

\section{Related Work}

Increasing attention has been paid to the security of semantic communications.
In \cite{du2023rethinking, shen2023secure}, the potential threats and secure requirements were discussed, and meanwhile,  the feasibility of possible defense mechanisms was analyzed under semantic communication scenarios.
The following section describes works on security of semantic communication systems from two specific perspectives: data integrity and privacy preservation.

\textit{Date Integrity of Semantic Communication System}: 
Targeted and non-targeted adversarial attacks with small perturbations was explored in \cite{sagduyu2023semantic} to manipulate the transmitted semantics.
In \cite{liu2023semprotector}, a semantic signature generation method is proposed based on generative adversarial networks to protect the integrity of semantics against adversarial perturbations over the end-to-end semantic communication system.
In distributed semantic communication systems, with a focus on efficient and secure information interaction in Web 3.0 and Metaverse, blockchain was introduced into semantic communications in \cite{lin2024blockchain,lin2023blockchain}. Tamper-resistant mechanisms inherent in blockchain and smart contracts were utilized to verify the integrity and authenticity of semantics, and validate the quality of semantics.
However, the current studies lack research on verifying the authenticity of lossy semantics.

\textit{Privacy Preservation of Semantic Communication System}:
A model inversion eavesdropping attack was proposed \cite{chen2023model} for semantic communications leading to leakage of private information, in which the attacker interpreted transmitted semantics within wireless channels and tried to reconstruct the original information by model inversion.
To resist the model inversion attack, a defense method based on random semantics permutation and substitution\cite{chen2023model} was proposed to prevent the attacker from efficiently reconstructing the original information.
By training encoders to maximize the reconstruction distortion of adversaries, the adversarial learning approach in \cite{wang2024privacy} was able to protect users’ privacy against model inversion attacks.
To address the privacy risk caused by knowledge discrepancies among communicating nodes, the knowledge discrepancy oriented privacy preserving method\cite{cheng2024knowledge} reduced the knowledge discrepancy between the sender and receiver by matching the unknown knowledge to the known prior knowledge.
In discrete task-oriented semantic communications, the adversarial learning was utilized to against information leakage\cite{zhang2024utility}.
However, current privacy-preserving schemes in semantic communications are limited to specific scenarios and tasks, and lack mathematically rigorous proof of privacy-preserving.

\section{System Model}
In a distributed IoT network, heterogeneous devices perform intelligent tasks with each other utilizing semantic communication codecs, and dynamically update their codecs in a distributed manner, as shown in Fig. \ref{fig_scenario_description}.
The whole process is comprised of two main phases, update and usage, which are detailed as follows:

\begin{figure}[t]
	\includegraphics[width=3.5in]{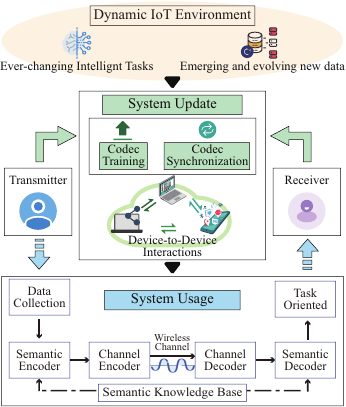}
	\caption{The semantic communication system in distributed IoT networks.}
	\label{fig_scenario_description}
\end{figure}
\begin{enumerate}
	\item{
		\textbf{Update:} To keep pace with the ever-changing demands of IoT tasks and emerging new data, the semantic communication system is not fixed and static, but is constantly trained and synchronized.
		\begin{enumerate}
			\item{
				\textbf{Training:} IoT devices collect evolving training data about tasks.
				Then, they perform federated learning to update the semantic communication system so that codecs can adapt to changing tasks and requirements.
			}
			\item{
				\textbf{Synchronization:} Since not all devices may participate in the training process because of limited resources, the synchronization phase is important to ensure that all devices receive the latest codecs. Furthermore, due to the inherent dynamic topology of IoT networks, where devices frequently join and leave the network, it is imperative for new IoT devices joining the network to retrieve the latest model to maintain consistency and coherence within the network. 
			}
		\end{enumerate}
	}
	\item{
		\textbf{Usage:} When synchronization is complete, IoT devices proceed to the usage phase, where they perform model inference using the latest codecs for task-oriented semantic communications with the collected data.
	}
\end{enumerate}
There are attackers in the scenario, categorized into internal and external attackers.
Internal attackers within IoT networks are ``honest and curious". They comply with network protocols, but out of curiosity or malicious intent, they conduct passive attacks, carrying out information eavesdropping or unauthorized analysis. For example, such an adversary attempts to perform model inversion attacks to gain access to sensitive data without disrupting system usage processes.
External attackers are from outside the IoT networks, and can launch active attacks in addition to passive attacks.
They initiate active attacks, including data tampering, data falsification, and denial-of-service attacks, with the aim of directly corrupting the update and usage processes.

\subsection{Semantic Communication System with Compressive Semantic Knowledge Base}
\label{semantic_communication_system_with_semantic_knowledge_base}

Without loss of generality, we concentrate on semantic communications for the text transmission task. 
The sentence with $E$ words to be transmitted in the semantic communication system is denoted as $\boldsymbol{s}=[w_1,w_2,\dots,w_E]$, where $w_e$ is the $e$-th word in the sentence.
The transmitter comprises three essential components: semantic encoder, channel encoder, and semantic knowledge base. 
The semantic encoder is responsible for transforming the input data into meaningful semantic features.
The semantic knowledge base provides the encoder with the fundamental understanding to improve the ability of semantic extraction.
The channel encoder, which follows the semantic encoder, converts and compresses the semantic representations into fewer signal symbols suitable for transmission over the communication channel, ensuring the reliable and efficient data delivery among IoT devices.
Specifically, the sentence is first embedded as $\boldsymbol{s}_{embed}\in \mathbb{R}^{E \times Q}$.
The transmitter then utilizes the semantic encoder to extract features from $\boldsymbol{s}_{embed}$ with the help of the knowledge base, denoted as
\begin{equation}
	\boldsymbol{f}= S_{\boldsymbol{\alpha}}\left( \boldsymbol{s}_{embed}|| \boldsymbol{\kappa}\right) ,
\end{equation}
where $\boldsymbol{\kappa}\in \mathbb{R}^{P \times Q}$ is a semantic knowledge base with $P$ vectors, each of size $Q$. $\boldsymbol{f}\in \mathbb{R}^{(S+P)\times Q}$ denotes extracted features. $S_{\boldsymbol{\alpha}}\left( \cdot\right) $ is the semantic encoder with the parameters $\boldsymbol{\alpha}$.
Afterward, the channel encoder processes $\boldsymbol{f}$ to obtain signal symbols to be transmitted $\boldsymbol{x}\in \mathbb{C}^{L\times1}$, represented as
\begin{equation}
	\label{eq_transmitter}
	\boldsymbol{x}=C_{\boldsymbol{\beta}}\left( \boldsymbol{f} \right) ,
\end{equation}
where $C_{\boldsymbol{\beta}}\left( \cdot\right) $ is the channel encoder with the parameters $\boldsymbol{\beta}$.
Taking into account the inevitable model noise, $\boldsymbol{x}$ is also represented as
	\begin{equation}
		\label{eq_x_si}
		\boldsymbol{x} = \boldsymbol{s_i} + \boldsymbol{n}_{model}, 
	\end{equation}
	where $\boldsymbol{s_i}$ is the semantic information accurately extracted from $\boldsymbol{s}$, and $\boldsymbol{n}_{model} \sim \mathcal{CN}\left( 0,\sigma_m^2\mathbf{I}_L\right) $ represents the model noise with Gaussian distribution, which is the result of unstable gradients descending, the training data noise and other factors\cite{xie2023semantic}.

The signal received at the receiver is
\begin{equation}
	\label{eq_sc_channel}
	\boldsymbol{y}=\boldsymbol{h}\boldsymbol{x}+\boldsymbol{n}_{channel},
\end{equation}
where $\boldsymbol{y}\in \mathbb{C}^{L\times1}$, $\mathbf{n}_{channel}$ is the additive white Gaussian noise (AWGN), following $\mathbf{n}_{channel} \sim \mathcal{CN}\left( 0, \sigma_n^2\mathbf{I}_L\right) $, $\boldsymbol{h}$ is the channel gain.
For the Rayleigh fading channel, $\boldsymbol{h}\sim\mathcal{CN}\left( 0,\mathbf{I}_L\right) $; and for Rician fading channel, $\boldsymbol{h}\sim\mathcal{CN}\left(\mu_h \mathbf{I}_{L\times1},\sigma_h^2 \mathbf{I}_L\right) $ with $\mu_h=\sqrt{r/(r+1)}$ and $\sigma_h=\sqrt{1/(r+1)}$, where $r$ is the Rician coefficient.
According to (\ref{eq_x_si}) and (\ref{eq_sc_channel}), the received signal can also be represented as
	\begin{equation}
		\label{eq_received_y}
		\boldsymbol{y}
		= \boldsymbol{h}\left( \boldsymbol{s_i} + \boldsymbol{n}_{model}\right)  + \boldsymbol{n}_{channel}.
\end{equation}

The receiver includes semantic decoder, channel decoder and semantic knowledge base. 
Its semantic knowledge base is synchronized to the transmitter's.
The channel decoder processes the received signals to recover semantic features, mitigating errors or distortions caused during the wireless communication process.
To be specific, the features recovered from $\boldsymbol{y}$ by the channel decoder is denoted as
\begin{equation}
	\hat{\boldsymbol{f}} = C^{-1}_{\boldsymbol{\psi}}\left( \boldsymbol{y}\right),
\end{equation}
where $C^{-1}_{\boldsymbol{\psi}}\left( \cdot\right)$ is the channel decoder with parameters $\boldsymbol{\psi}$.
Subsequently, the semantic decoder leverages the semantic knowledge base to decode these features, represented as
\begin{equation}
	\hat{\boldsymbol{s}}=S_{\boldsymbol{\chi}}^{-1}\left( \hat{\boldsymbol{f}}||\boldsymbol{\kappa}\right) ,
\end{equation}
where $\hat{\boldsymbol{s}}$ is the recovered sentence, and $S_{\boldsymbol{\chi}}^{-1}\left( \cdot\right) $ is the semantic decoder with parameters $\boldsymbol{\chi}$.

\subsection{Problem Definition}
\subsubsection{Designing an Efficient Update Scheme with Compressive Semantic Knowledge Bases}

The substantial volume of data exchange required during the process of collaborative training and synchronizing $\boldsymbol{\alpha}$, $\boldsymbol{\chi}$, $\boldsymbol{\beta}$, $\boldsymbol{\psi}$ and $\boldsymbol{\kappa}$ poses a challenge of updating efficiency.
Updating only the compressive knowledge base is expected to solve this challenge.
This requires refining semantic knowledge bases to achieve a small number of vectors, $P$, while maintaining their semantic richness.
The refinement is crucial for reducing transmission overheads during system update and empowering the semantic codec with the fundamental knowledge.
\subsubsection{Achieving a Flexible Semantic Coding Method for Heterogeneous Devices}
The wide range of transmission and computation capabilities requires the system to be flexible.
In the system usage phase, the transmission capability restricts the maximum value of the transmitted signal length $L$, and the computation capability limits the number of semantic knowledge vectors $P$ involved in model inference. 
The goal of the flexible semantic coding method can be represented as
\begin{equation}
	\max \zeta_{L,P}\left( \boldsymbol{s}, \hat{\boldsymbol{s}}\right) \quad\forall L\in\boldsymbol{L}, P\in\boldsymbol{P}\\,
\end{equation}
where $\boldsymbol{L}$ represents the set of numbers of symbols that devices can transmit, and $\boldsymbol{P}$ is the set of numbers of semantic knowledge vectors that devices can use, $\zeta_{L,P}\left( \cdot,\cdot\right) $ measures the similarity between $\boldsymbol{s}$ and $\hat{\boldsymbol{s}}$ when device transmits $L$ symbols and utilize $P$ semantic communication vectors.
In the update phase, the larger $L$ and $P$ means the more computation burden in models training. Therefore, the flexible of the system update process refers to allowing heterogeneous devices to select $L$ and $P$ based on their own computing capability during model training.
\subsubsection{Verifying the Integrity and Authenticity of Transmitted Semantics}
The nature of lossy transmission in semantic communications determines that $\boldsymbol{f}$ and $\hat{\boldsymbol{f}}$ are not the same, because transmitted semantics is inevitably affected by model noise and the wireless channel.
This leads to the unavailability of the traditional signature method and poses a serious challenge in verifying the integrity and authenticity of the semantics.
The verification mechanism is required to check for the semantics manipulation and falsification with tolerance to the effects of $\boldsymbol{n}_{model}$, $\boldsymbol{n}_{channel}$ and $\boldsymbol{h}$.

\subsubsection{Providing Transparent Differential Privacy in Semantic Communications}
Considering potential privacy leakages during system usage phase, we need to design a differential privacy mechanism for semantic communications.
By adding differential privacy noise to the transmitted signal symbols, the DP mechanism can effectively prevent attackers from performing malicious data analysis.
However, in semantic communications, the transmitted semantics are also affected by model noise and wireless channel noise.
These noises also contribute a certain level of differential privacy protection.
It requires a DP mechanism with joint analyses of these three types of noise, to achieve data privacy preservation with the least added noise.

\section{Proposed Solution}

To ensure the integrity and availability of the update process in the distributed semantic communication system, we propose a blockchain-based trustworthy update scheme.
Based on above trusted scheme, an efficient and flexible semantic coding method is designed to implement system updates with fewer data exchanges and to provide flexibility for heterogeneous devices.
In the system usage phase, we introduce a signature mechanism for lossy semantics to guarantee the integrity and authenticity of semantics transmitted over lossy channels.
In response to privacy leakage threats in task-oriented semantic communications, a noise-aware differential privacy mechanism is proposed to defend against data analysis attacks.

\subsection{Blockchain-based Trustworthy Update Scheme}\label{contribution_1}
\begin{figure}[t]
	\includegraphics[width=3.5in]{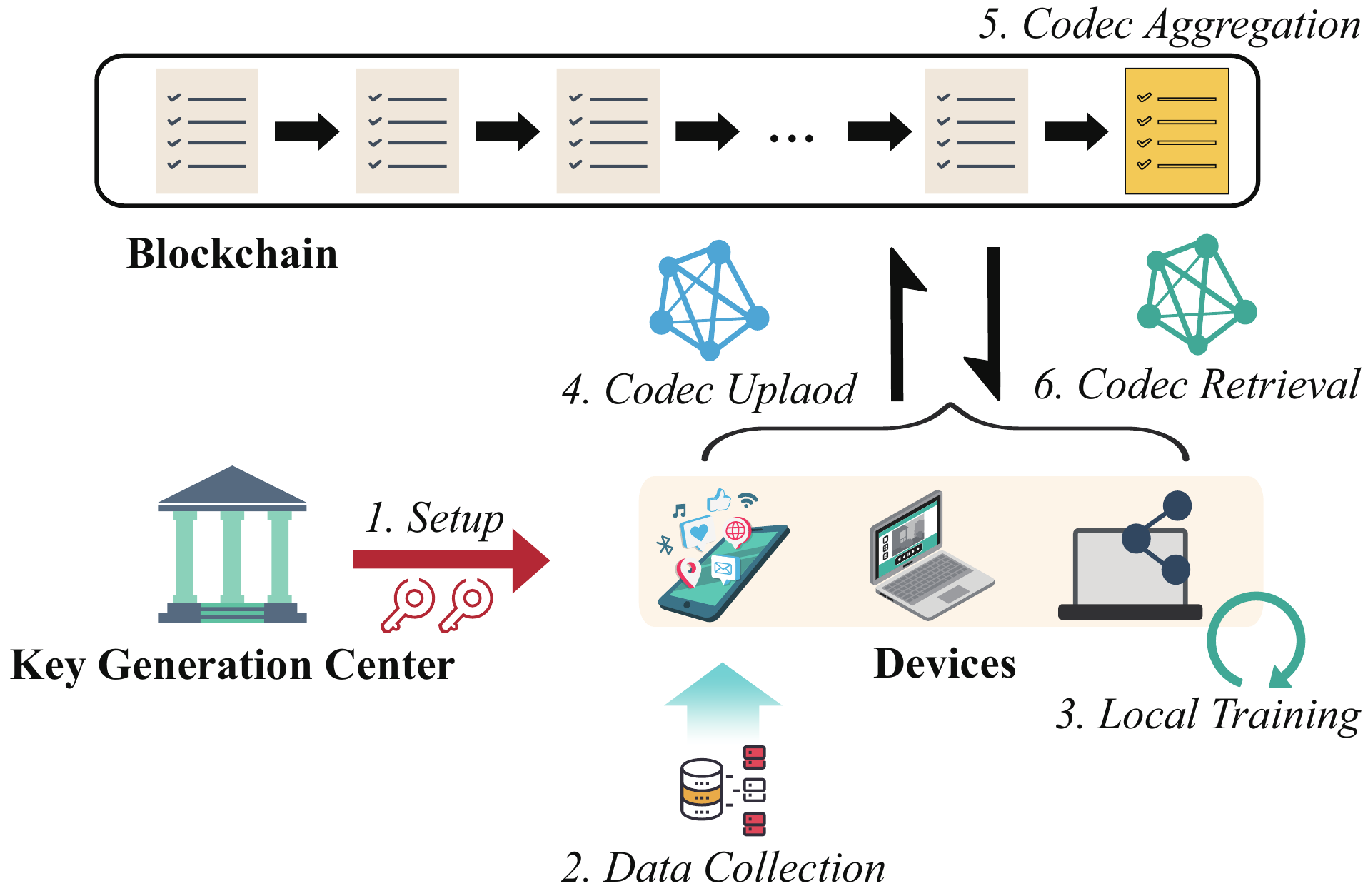}
	\caption{The workflow of blockchain-based trustworthy update scheme.}
	\label{fig_overview}
\end{figure}

We design a trustworthy scheme based on blockchain for the integrity and availability of the system update. Its workflow is shown in the Fig. \ref{fig_overview}.
The scheme consists of three entities, which are elaborated as follows:
\begin{enumerate}
	\item{
		\textbf{IoT devices:} 
		They have initial semantic communication codecs for performing tasks-oriented communications, and interact with each other to continuously update and synchronize the distributed semantic communication system.
		In addition, there are error-free links between them through protocols such as Bluetooth or WiFi that have been widely integrated into the IoT ecosystem.
	}
	\item{
		\textbf{Key Generation Center:} A trusted third party plays a crucial role in the network, facilitating network initiation and public/private key pairs generation and distribution\cite{miao2022privacy}. It is worth noting that the center is unable to directly organize interactions and perform complicated data processing, due to availability issues caused by complex IoT environments or the limited computing and communication capabilities of the center.
	}
	\item{
		\textbf{Blockchain:} A consortium blockchain\cite{belotti2019vademecum, androulaki2018hyperledger} is a intangible, conceptual entity maintained by IoT devices. 
		It is a distributed immutable ledger, constructed as a list of blocks. Each block records a set of transactions, where a transaction represents an operation to read or write data to the ledger. 
		It records all processes of system training and synchronization as transactions.
		Each device maintains a copy of the ledger by a collaborative process called consensus, ensuring the proper execution of operations, the validation of blocks, and the consistency of the ledger among peers.
		The blockchain is fault-tolerant and can withstand single point of failures.
		
	}
\end{enumerate}

The workflow of trustworthy update scheme consists of the following steps.
\begin{enumerate}
	\item{
		\textit{Setup:} IoT devices register in the key generation center, where they obtain pairs of public and privacy keys, denoted as $\left( pk,sk\right) $.
	}
	\item{
		\textit{Data Collection:} IoT devices collect evolving data relevant to the semantic communication tasks.
	}
	\item{
		\textit{Local Training:} Utilizing the collected data, IoT devices train their local models or semantic knowledge bases, which will be described in detail in Section \ref{section_cskb}.
	}
	\item{
		\textit{Codec Upload:} Devices upload parameters of locally trained codecs to the blockchain.
	}
	\item{
		\textit{Codec Aggregation:} It is executed on the blockchain, and supports the use of any federated learning solutions. We select a simple and widely used federated algorithm, FedAvg\cite{mcmahan2017communication}, to generate the updated codec.
	}
	\item{
		\textit{Codec Retrieval:} Devices retrieve up-to-date codecs from blockchain to synchronize codecs across the network.
	}
\end{enumerate}

\subsection{Efficient and Flexible Semantic Coding Method based on Compressive Semantic Knowledge Base}\label{section_cskb}
To solves the challenge of system inefficiency caused by transmitting large amounts of data during system update, we propose a semantic coding method to realize the system update by only training and synchronizing the compressive knowledge base.
Moreover, a forward propagation with the pruning mechanism is designed for heterogeneous IoT devices, achieving the adjustment of the size of semantic knowledge base and transmitted symbols according to their transmission and computation resources in model training and inference stages.

Consisting of multiple semantic knowledge vectors, the compressive semantic knowledge base is the core of the method.
They provide the semantic codec with task-relevant background knowledge in the usage phase to achieve superior communication performance.
In addition, during the update phase, the system is able to only synchronize and train the semantic knowledge base, enabling efficient maintenance with low data exchange.
Considering the diversity of semantic communication tasks, each task has a list of semantic knowledge vectors tailored specifically for it.
We define a list of semantic knowledge vectors for the semantic communication task $t$ as $\boldsymbol{\kappa}^t=\left[  \boldsymbol{v}_1^t, \boldsymbol{v}_2^t,\cdots, \boldsymbol{v}_{P^t}^t\right]  $, where $P^t$ is the total number of vectors, and $\boldsymbol{v}_p^t \in \mathbb{R}^{Q}$ represents the $p$-th $Q$-dimensional vector in $\boldsymbol{\kappa}^t$.
The semantic knowledge vectors are generated by a neural network, called as semantic knowledge network, with fixed inputs $\boldsymbol{FI}$.
We denote this network with parameters $\boldsymbol{\omega}$ by $K_{\boldsymbol{\omega}}$.
It is used to update semantic knowledge vectors during the training process. In the usage phase, devices can use its output directly without model inference with $K_{\boldsymbol{\omega}}$.

To achieve that heterogeneous devices flexibly adjust the computation and transmission overheads of the model training and model inference, we propose a forward propagation with the pruning mechanism to train the codec.
It supports adjusting the size of $\boldsymbol{\kappa}^t$ and $\boldsymbol{f}$, shown in Algorithm \ref{al_forward_propagation}.
Let $\boldsymbol{\kappa}^t_\varsigma$ represent a subsequence of $\boldsymbol{\kappa}^t$ comprising the first $\varsigma$ elements, and $\boldsymbol{f}_\varpi$ denote a subsequence of $\boldsymbol{f}$ containing the first $\varpi$ elements.
For each batch during training, $\varsigma$ and $\varpi$ are selected, ranging from one to the largest $\varsigma$ and $\varpi$ acceptable to the device, denoted as $\varsigma_{max}$ and $\varpi_{max}$.

\begin{algorithm}[]  
	
	\caption{Forward propagation with the pruning mechanism}
	\label{al_forward_propagation}
	\KwIn{batch data $\boldsymbol{S}$ from $D$, $\varsigma$, $\varpi$\;}
	$K_{\boldsymbol{\omega}}\left( \boldsymbol{FI}\right) \to \boldsymbol{\kappa}^t$\;
	\textbf{Transmitter:}

	\quad $S_{\boldsymbol{\alpha}}\left( \boldsymbol{S}|| \boldsymbol{\kappa}^t_\varsigma\right) \to \boldsymbol{f}$\;
	\quad Transmit $\boldsymbol{f}_\varpi$ over the channel\;
	
	\textbf{Receiver:}
	
	\quad Receive $\hat{\boldsymbol{f}_\varpi}$\;
	\quad $S_{\boldsymbol{\chi}}^{-1}\left( \hat{\boldsymbol{f}_\varpi}|| \boldsymbol{\kappa}^t_\varsigma\right) \to \hat{\boldsymbol{S}}$\;
	
	\KwOut{$\boldsymbol{f}$, $\hat{\boldsymbol{f}}$, $\hat{\boldsymbol{S}}$ }	
\end{algorithm}
Based on above forward propagation, the training of the semantic communication system is divided into three steps, the individual training of the semantic codec, the semantic knowledge base and the overall training of the whole system, as exhibited in Algorithm \ref{al_local_updates}.
In the first step, $S_{\boldsymbol{\alpha}}$ and $S_{\boldsymbol{\chi}}^{-1}$ are updated with the goal of minimizing the divergence between $\boldsymbol{s}$ and $\hat{\boldsymbol{s}}$. 
To quantify this divergence, we employ the cross-entropy (CE) to quantify the divergence, which is given by
\begin{equation}
	\label{eq_loss_ce}
	\begin{aligned}
		&\mathcal{L}_{CE}
		\left( \boldsymbol{s}, \hat{\boldsymbol{s}}\right) =\\
		&-\sum_{e=1}q\left( w_e\right) log\left( p\left( w_e\right) \right) + 
		\left( 1-q\left( w_e\right) \right) log\left( 1-p\left( w_e\right) \right) ,
	\end{aligned}
\end{equation}
where $q\left( w_e\right) $ denotes the real probability of the occurrence of $w_e$ in the original sentence $\mathbf{s}$, and $p\left( w_e\right) $ is the predicted probability of the same $w_e$ appearing in the  reconstructed sentence $\hat{\mathbf{s}}$. 
In the second step, $\boldsymbol{\kappa}^t$ is also updated with the goal of minimizing  $\mathcal{L}_{CE}$.
Furthermore, to ensure the broad representational capability of the semantic knowledge base and to make each vector with different knowledge, the conditional information entropy between vectors need to be maximized, represented as
\begin{equation}
 \max_{\boldsymbol{\kappa}^t}  H(\boldsymbol{\kappa}^t_{i_{1}}|\boldsymbol{\kappa}^t_{i_{2}}),\quad \forall 1\leq i_1,i_2\leq \varsigma_{max},i_1\neq i_2.
\end{equation}
Considering that the conditional entropy is minimum when two distributions agree, we set the training objective to minimize the cosine similarity between vectors in order to maximize the difference between their distributions\cite{xu2023q}.
By combining the above two optimization objectives, the $\mathcal{L}_{\boldsymbol{\kappa}}$ is set to
\begin{equation}
	\label{eq_loss_kappa}
	\mathcal{L}_{\kappa} \left( \boldsymbol{s}, \hat{\boldsymbol{s}}\right) =\lambda_1 \mathcal{L}_{CE} \left( \boldsymbol{s}, \hat{\boldsymbol{s}}\right) + \lambda_2\left|\left| \left( \boldsymbol{\kappa}^t\right)^{T}\left( \boldsymbol{\kappa}^t\right)\ \right|  \right|_2  ,
\end{equation}
where $\lambda_1,\lambda_2$ are weights to balance the loss.
\begin{algorithm}[]  
	
	\caption{Local update for semantic communication codec}
	\label{al_local_updates}
	
	\SetKwFunction{FunctionTraintheSemanticCodec}{Train the Semantic Codec}
	\SetKwProg{Fn}{Function}{:}{}
	\Fn{\FunctionTraintheSemanticCodec{}}{
		\KwIn{batch data $\boldsymbol{S}$ from dataset\;}
		
		Freeze $C_{\boldsymbol{\beta}}$, $C_{\boldsymbol{\psi}}^{-1}$, $\boldsymbol{\kappa}^t$\;
		Forward propagation based on Algorithm \ref{al_forward_propagation}\;
		Compute loss function $\mathcal{L}_{CE}$ by (\ref{eq_loss_ce})\;
		Train $S_{\boldsymbol{\alpha}}$, $S_{\boldsymbol{\chi}}^{-1}$ $\to$ Gradient descent with $\mathcal{L}_{CE}$\;
		\KwOut{$S_{\boldsymbol{\alpha}}$, $S_{\boldsymbol{\chi}}^{-1}$\;}
	}

	\SetKwFunction{FunctionTraintheSemanticKnowledgeBase}{Train the Semantic Knowledge Base}
	\SetKwProg{Fn}{Function}{:}{}
	\Fn{\FunctionTraintheSemanticKnowledgeBase{}}{
		\KwIn{batch data $\boldsymbol{S}$ from dataset\;}
		
		Freeze $C_{\boldsymbol{\beta}}$, $C_{\boldsymbol{\psi}}^{-1}$, $S_{\boldsymbol{\alpha}}$, $S_{\boldsymbol{\chi}}^{-1}$\;
		Forward propagation based on Algorithm \ref{al_forward_propagation}\;
		Compute loss function $\mathcal{L}_{\kappa}$ by (\ref{eq_loss_kappa})\;
		Train $\boldsymbol{\kappa}^t$ $\to$ Gradient descent with $\mathcal{L}_{\kappa}$\;
		\KwOut{$\boldsymbol{\kappa}^t$\;}
	}
	\SetKwFunction{FunctionTraintheWholeSystem}{Train the Whole System}
	\SetKwProg{Fn}{Function}{:}{}
	\Fn{\FunctionTraintheWholeSystem{}}{
		\KwIn{batch data $\boldsymbol{S}$ from dataset\;}
		Forward propagation based on Algorithm \ref{al_forward_propagation}\;
		Compute loss function $\mathcal{L}_{\kappa}$ by (\ref{eq_loss_kappa})\;
		Train $S_{\boldsymbol{\alpha}}$, $S_{\boldsymbol{\chi}}^{-1}$, $C_{\boldsymbol{\beta}}$, $C_{\boldsymbol{\psi}}^{-1}$, $\boldsymbol{\kappa}^t$ $\to$ Gradient descent with $\mathcal{L}_{\kappa}$\;
		\KwOut{$S_{\boldsymbol{\alpha}}$, $S_{\boldsymbol{\chi}}^{-1}$, $C_{\boldsymbol{\beta}}$, $C_{\boldsymbol{\psi}}^{-1}$, $\boldsymbol{\kappa}^t$\;}	
	}
\end{algorithm}

The system initialization can perform all functions in Algorithm \ref{al_local_updates}, while in the system update phase, IoT devices only train the semantic knowledge base based on (\ref{eq_loss_kappa}) and synchronize it with less data exchange for efficient updates.
Furthermore, devices have ability to balance communication performance with transmission and computation costs by flexibly pruning $\boldsymbol{\kappa}^t$ and $\boldsymbol{f}$.
In the model inference stage, devices with limited computing power can negotiate with each other to truncate the $\boldsymbol{\kappa}^t$ for mitigating the computational cost of the semantic encoding and decoding processes. 
The transmitter can also trim $\boldsymbol{f}$ to reduce the number of signal symbols to be transmitted.
For the model training, devices are able to decrease $\varsigma_{max}$ and $\varpi_{max}$ to reduce the training computational consumption.

\subsection{Signature Mechanism for Lossy Semantics}
Transmitting semantics over open wireless channels faces the challenge of protecting its integrity and authenticity. Although cryptographic mechanisms, such as digital signatures, are widely used to address this challenge, they cannot be directly applied in semantic communications. 
This is because the lossy nature of semantic communications would inevitably lead to the failure of signature verification. Therefore, we design a signature mechanism for lossy semantics to achieve secure semantic communications. To avoid the model noise from affecting integrity verification, the mechanism is applied at the signal symbol level.

\begin{figure}[t]
	\centering
	\includegraphics[width=3.5in]{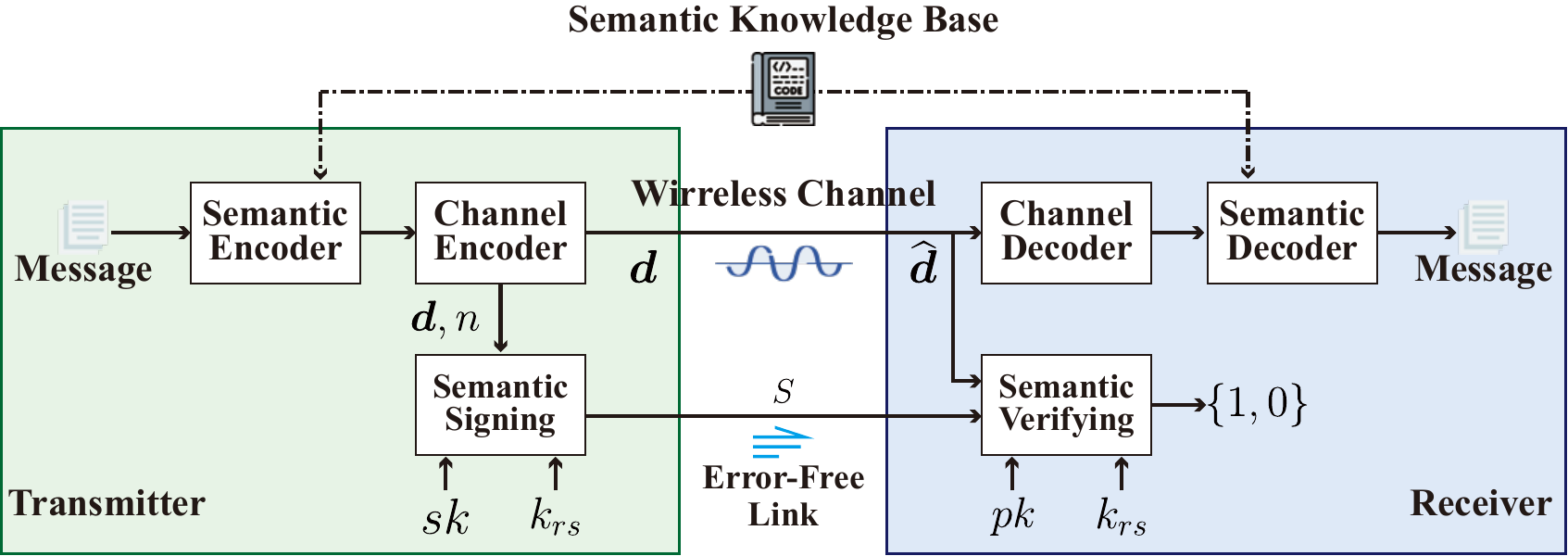}
	\caption{Illustration of the signature mechanism for integrity and authenticity of lossy semantics.}
	\label{fig_signature mechanism}
\end{figure}

The signature mechanism is shown in the Fig. \ref{fig_signature mechanism}.
The output of the channel encoder is split into blocks.
We denote the $n$-th block as $\boldsymbol{d}\triangleq \{d_i\in \mathbb{R}^m|i=1,...,I\}$.
The block goes through the wireless channel, and is received by receiver, denoted as $\widehat{\boldsymbol{d}}$.
The signature mechanism contains two algorithms: $SemanticsSigning$ and  $SemanticsVerifying$. $SemanticsSigning$ takes $\boldsymbol{d}$ and the privacy key of transmitter $sk$ as inputs and outputs the signature of $\boldsymbol{d}$, denoted as $S$.
Then $S$ is transmitted to the receiver via error-free links.
The receiver executes the Algorithm $SemanticsVerifying$ with inputs $\widehat{\boldsymbol{d}}$, $S$ and the transmitter's public key $pk$, verifying the integrity and source of $\widehat{\boldsymbol{d}}$.

\begin{figure}[t]
	\centering
	\includegraphics[width=3.5in]{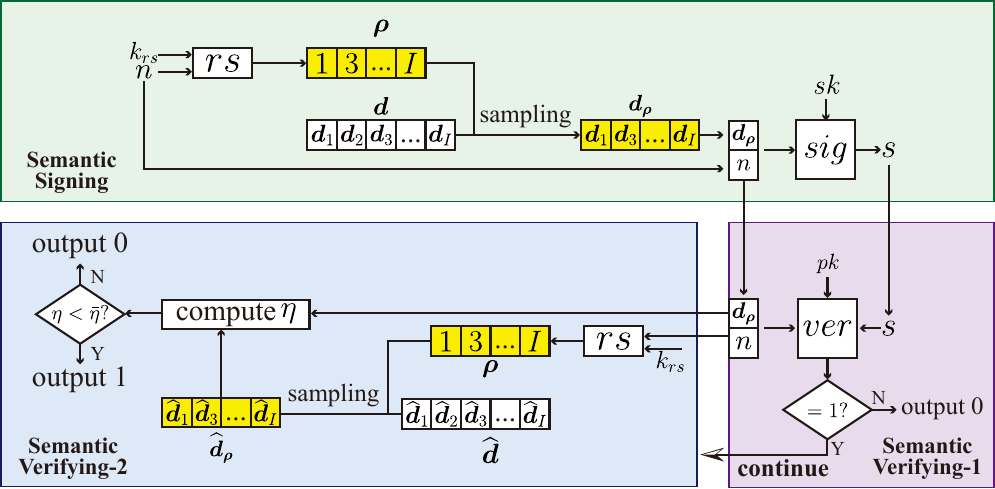}
	\caption{The workflow of the proposed signature mechanism for lossy semantics.}
	\label{fig_sig_detial}
\end{figure}

The detailed explanation of the mechanism is as follows.
\begin{enumerate}
	\item{
		\textit{Setup:} Receiver obtain $pk$ of the transmitter from the key generation center.
		The two communicating parties and the key generation center negotiate relevant parameters of the following functions:
		\begin{enumerate}
			\item{
				$sig(me,sk)\to s$ and $ver(me,s,pk)\to \{0,1\}$: Signature generation and verification functions of a standard digital signature scheme, such as Digital Signature Algorithm (DSA)\cite{pub2000digital}.
				$(pk,sk)$ is a public-private key pair of the transmitter, and $me$ is the message to be signed.
			}
			\item{
				$rs(a,k_{rs})\to \boldsymbol{\rho}$: A pseudorandom function, where $k_{rs}$ is a key shared by the two parties, $a$ is an arbitrary input, $\boldsymbol{\rho}$ is a random index set with $|\boldsymbol{\rho}|$ non-repeating integers between $1$ and $I$.
			}
		\end{enumerate}
	}
	\item{
		$SemanticsSigning\left(\boldsymbol{d},n,sk,k_{rs}\right) \to S$:
		As shown in the upper part of Fig. \ref{fig_sig_detial}, this algorithm first gets random index set by computing $\boldsymbol{\rho} = rs(n,k_{rs})$.
		After that, $\boldsymbol{d}$ is sampled based on $\boldsymbol{\rho}$, getting result denoted as $\boldsymbol{d}_{\boldsymbol{\rho}}\triangleq\left\lbrace d_{i}|i\in\boldsymbol{\rho}\right\rbrace $.
		Transmitter signs $\boldsymbol{d}_{\boldsymbol{\rho}}$ and the index of block $n$ with its privacy key $sk$, represented as $s = sig(\{\boldsymbol{d}_{\boldsymbol{\rho}}||n\},sk)$.
		The output of this algorithm is defined as $S\triangleq\left\lbrace {\boldsymbol{d}}_{\boldsymbol{\rho}}  ||n||s \right\rbrace$.
		
	}
	\item{
		$SemanticsVerifying(\widehat{\boldsymbol{d}},S,pk,k_{rs}) \to \{1,0\}$: 
		There are two steps in the algorithm, as illustrated in the lower part of Fig. \ref{fig_sig_detial}.
		First, this algorithm gets $s$ and $ \{\boldsymbol{d}_{\boldsymbol{\rho}}  ||n\} $ from $S$ and compute $ver(\{\boldsymbol{d}_{\boldsymbol{\rho}}||n\},s,pk)$.
		An output of $1$ indicates that $\{\boldsymbol{d}_{\boldsymbol{\rho}}||n\}$ indeed originates from the transmitter and has not been tampered with.
		A result of $0$ means that the verification of $ver$ is failed and this algorithm also returns $0$.
		After the signature is validated, $n$ is checked for its freshness to defend against replay attacks.
		The algorithm then generates $\boldsymbol{\rho}$ by calculating  $rs(n,k_{rs})$ and samples $\widehat{\boldsymbol{d}}$ based on $\boldsymbol{\rho}$, getting ${\widehat{\boldsymbol{d}}}_{\boldsymbol{\rho}}\triangleq\{ \widehat{{d}}_{i}|i\in\boldsymbol{\rho}\} $.
		Finally, the difference between ${{\boldsymbol{d}}}_{\boldsymbol{\rho}}$ and ${\widehat{\boldsymbol{d}}}_{\boldsymbol{\rho}}$ is evaluated as $\eta\triangleq \lbrace \| d_{i}-\widehat{{d}}_{i}\| _2=\eta_i|i\in\boldsymbol{\rho}\rbrace$. The difference is compared to the specified threshold $\bar \eta$, which is described in detail below. The algorithm returns $1$ if the validation passes, otherwise it returns $0$. The algorithm with a time complexity of $O(|\boldsymbol{\rho}|)$ dose not impose a serious computational burden on devices.
	}
	
\end{enumerate}

To ensure the completeness of the signature mechanism, which means that the mechanism always passes validations in the absence of attackers, the effect of channel noise on the data is taken into account in the design of $\bar \eta$.
We discuss the design based on the classification of semantic communications into utilizing finite constellation and full-resolution constellation. Our mechanism has a good compatibility.
\begin{enumerate}
	\item{
		For semantic communications with finite constellation, $d_i$ is configured as $\mathbb{R}^1$ to describe the real or imaginary parts of signal symbols. The threshold $\bar\eta$ is set to half of the distance between points in the constellation. The validation passes if $\bar\eta > \eta_i$, which means that received symbol related to $d_i$ is mapped to the right constellation point, otherwise it means that the mapping is to the wrong constellation point.
	}
	\item{
		In semantic communications with full-resolution constellation, $d_i$ represents constellation points of the latent semantic codewords.
		We map $d_i$ into the $m$-dimension sphere space as a point, which has a noise sphere with radius $r_c\triangleq\sqrt{m}\sigma_{channel}$ in AWGN\cite{xie2023semantic}.
		For semantic communications with non-overlapping noise sphere, the point of $\widehat{\boldsymbol{d}}$ mapped into the $m$-dimension sphere space is within this noise sphere.
		Therefore, the threshold $\bar \eta$ is set to ${r_c}$.
		A $\eta_i$ less than $\bar\eta$ means that the interference with $\widehat{{d}}_{i}$ during the transmission is within the normal interval, otherwise  $\widehat{{d}}_{i}$ has been tampered with.
		
	}

\end{enumerate}

For the successful verification of data integrity, adversaries can not know which elements of $\boldsymbol{d}$ will be sampled  until the transmission of $\boldsymbol{d}$ is complete in semantic communications.
Once adversaries are aware of it before receiver receives $\boldsymbol{d}$, they are able to launch attacks without being detected by only modifying the data whose index is not in $\boldsymbol{\rho}$.
Therefore, it is crucial to ensure that the random sampling key $k_{rs}$ is not leaked, and $S$ must be transmitted after $\boldsymbol{d}$ or encrypted.

In order to analyze the security of this mechanism, we first classify attacks on this signature mechanism into two categories, based on whether $\eta$ after the attack is greater than the predefined $\bar{\eta}$.
For attacks where $\eta\geq\bar{\eta}$, the detection probability will increase as the size of $\boldsymbol{\rho}$ increases.
When $x(x\leq I)$ items are modified in $\boldsymbol{d}$, the probability of detection is
\begin{equation}
	P_d = 1-\frac{C_{I-x}^{|\boldsymbol{\rho}|}}{C_{I}^{|\boldsymbol{\rho}|}}.
\end{equation}
The mechanism is also resistant to attacks when $\eta<\bar{\eta}$, such as adversarial attacks, which introduce small artificial noise into $\boldsymbol{d}$, causing the model to make incorrect predictions.
The design of $\bar \eta$ is based on the upper bound of the impact of channel noise on $\boldsymbol{d}$, which greatly limits the level of malicious artificial noise can be added without being detected.
Hence, the small artificial noise is overwhelmed by channel noise, model noises and differential privacy noise mentioned in Section \ref{section_dp}, hardly deteriorating model predictions\cite{wang2023adversarial}.

\subsection{The Noise-Aware Differential Privacy Mechanism}\label{section_dp}
To preserve the data privacy with less performance loss in the semantic communication system, we propose a noise-aware differential privacy mechanism that reduces the additional noise that needs to be introduced by jointly considering channel effects and model noise.
The mechanism works at the signal symbol level, so it can uniformly and transparently provide differential privacy for any communication tasks.

Formally, a function $\mathcal{F}:\boldsymbol{D}\to \boldsymbol{Y}$ satisfies $\left( \epsilon,\delta\right) $-differential privacy\cite{dwork2006calibrating,xue2023differentially} if and only if for any two adjacent datasets $\mathcal{D},\mathcal{D}'\subseteq \boldsymbol{D}$ and outputs $\boldsymbol{\gamma}\subset\boldsymbol{Y}$, we have
\begin{equation}
	\label{eq_dp}
	Pr[\mathcal{F}(\mathcal{D}) \in \boldsymbol{\gamma}] \leq e^{\epsilon} Pr[\mathcal{F}(\mathcal{D}') \in \boldsymbol{\gamma}] + \delta,
\end{equation}
where $\mathcal{D}$ and $\mathcal{D}'$ differ in only one sample, $\boldsymbol{D}$ and $\boldsymbol{Y}$ are sets of all dataset $\mathcal{D}$ and output $\boldsymbol{y}$ respectively, 
$\epsilon$ controls the privacy loss, with smaller values indicating stronger privacy protection, 
and $\delta$ allows for a small probability of deviation from the strict privacy guarantee, providing a flexible approach in scenarios where absolute privacy may be impractical.
Hence, a mechanism satisfies $\left( \epsilon,\delta\right) $-differential privacy if, for any pair of adjacent datasets, and for any  outputs, the ratio of the probabilities of observing these outputs under the mechanism is bounded by $e^{\epsilon}$ with probability at least $1 - \delta$.
Note that $\triangle$ is sensitivity of the function, defined as the maximum of $||\mathcal{F}(\mathcal{D})-\mathcal{F}(\mathcal{D}')||_2$.

Considering that the the signal symbol is in complex domain, it is necessary to extend the existing differential privacy mechanism to the complex domain.
We propose a complex Gaussian difference privacy mechanism following \cite{balle2018improving}. Specifically, for function $f:\boldsymbol{D}\to \mathbb{C}^{d}$ with sensitivity $\triangle$, $f(\mathcal{D})+Z$ with $Z\in \mathcal{CN}(0,2\sigma^2I)$ is $\left( \epsilon,\delta\right) $-differential privacy if $\sigma$ is calculated based on Algorithm \ref{al_complex_gaussian_dp}, where $\Phi$ is the cumulative density function of the standard univariate Gaussian distribution.
\begin{algorithm}[]  
	
	\caption{Computing $\sigma$ in Complex Gaussian Difference Privacy Mechanism}
	\label{al_complex_gaussian_dp}
	\KwIn{$\triangle$, $\epsilon$, $\delta$\;}
	
	$\Phi\left(0\right)-e^{\epsilon}\Phi\left(-\sqrt{2\epsilon}\right)   \to \delta_0$\;
	
	\If {$\delta \geq \delta_0$}  {
		\textbf{Define} $B_{\epsilon}^{+}\left( v\right) =\Phi\left( \sqrt{\epsilon v}\right) -e^\epsilon\Phi(-\sqrt{\epsilon(v+2)} ) $\;
		$\sup\left\lbrace v\in\mathbb{R}_{\geq 0}:B_{\epsilon}^{+}\left( v\right)\leq\delta\right\rbrace \to v^*$\;
		$\sqrt{1+v^*/2}-\sqrt{v^*/2 }\to \alpha$\;
	}
	
	\Else  {
		\textbf{Define} $B_{\epsilon}^{-}\left( u\right) =\Phi\left( -\sqrt{\epsilon u}\right) -e^\epsilon\Phi(-\sqrt{\epsilon(u+2)} ) $\;
		$\inf\left\lbrace u\in\mathbb{R}_{\geq 0}:B_{\epsilon}^{-}\left( u\right)\leq\delta\right\rbrace \to u^*$\;
		$\sqrt{1+u^*/2}+\sqrt{u^*/2 }\to \alpha$\;
	}
	$\alpha\triangle/\sqrt{2\epsilon}\to\sigma$\;
	
	\KwOut{$\sigma$ }	
\end{algorithm}

With Algorithm \ref{al_complex_gaussian_dp}, we can compute the variance of the target noise that needs to be added, denoted as $\sigma_t^2$, for the target $(\epsilon_t,\delta_t)$-differential privacy.
The differential privacy has post-processing immunity property, which guarantees that any additional computation or analysis performed on the output of a differential private algorithm does not compromise its privacy guarantees\cite{zhu2021bias}.
Therefore, we can introduce the target noise at any stage in semantic communications before the attacker receives the signal, and the target noise can be contributed by a combination of multiple noises.

To be specific, we add differential privacy noise, defined as $\mathbf{n}_{dp} \sim \mathcal{CN}( 0, \sigma_{dp}^2\mathbf{I}_L) $ to the signal to be transmitted $\boldsymbol{x}$.
Therefore, based on (\ref{eq_received_y}), the received signal is
\begin{equation}
	\label{eq_y_with_dp}
	\boldsymbol{y}
	= \boldsymbol{h}\left( \boldsymbol{s_i} + \boldsymbol{n}_{model} + \boldsymbol{n}_{dp}\right)  + \boldsymbol{n}_{channel}.
\end{equation}
Because the model noise and channel noise are immutable, we adjust the differential privacy noise to achieve the target with minimum additional noise.
The following discussion of determining $\sigma_{dp}^2$ is based on whether the IoT device can estimate $\boldsymbol{h}$, which fully accounts for the variability of the noise estimation capability of each device.

With a given $\boldsymbol{h}$, $\boldsymbol{y}$ follows $\mathcal{CN}\left(  \boldsymbol{h}\boldsymbol{s_i}, \sigma_{j}^2\boldsymbol{I}\right) $, where 
\begin{equation}
	\sigma_{j}^2 = \left|\boldsymbol{h} \right|^2\left( \sigma_{model}^2+\sigma_{dp}^2\right) +\sigma_{channel}^2.
\end{equation}
To make sure that $\sigma_{j}^2 > \sigma_{t}^2$ and thus achieve the target $(\epsilon_t,\delta_t)$-differential privacy, $\sigma_{dp}^2$ is set to
\begin{equation}
	\label{eq_with_h}
	\sigma_{dp}^2 = \max
	\left\lbrace 
	\left( \sigma_{t}^2-z_{c}\sigma_{channel}^2\right) / \left|\boldsymbol{h} \right|^2 - z_{m}\sigma_{model}^2
	,
	0
	\right\rbrace ,
\end{equation}
where $z_{c}$ and $z_{m}$ are binary numbers. The value of $1$ indicates that the IoT device is capable of measuring $\sigma_{channel}$ and $\sigma_{model}$, respectively, and a value of $0$ indicates that they cannot.

When $\boldsymbol{h}$ is unknown, the distribution of $\boldsymbol{y}$ is difficult to estimate.
To address this challenge, we consider $\boldsymbol{n}_{dp}$, $\boldsymbol{n}_{model}$ and $\boldsymbol{n}_{channel}$ independently. These noises provide  $(\epsilon_{dp},\delta_{dp})$, $(\epsilon_{model},\delta_{model})$, $(\epsilon_{channel},\delta_{channel})$-differential privacy, respectively. The post-processing immunity property of differential privacy means that as long as one of noises is larger than the target noise, this communication achieves differential privacy. Thus $\sigma_{dp}^2$ is determined as
\begin{equation}
	\label{eq_without_h}
		\sigma_{dp}^2 =
		\left\lbrace 
		\begin{aligned}
		0\quad&\max\left\lbrace z_{m}\sigma_{model}^2,z_{c}\sigma_{channel}^2\right\rbrace \geq \sigma_{t}^2\\
		\sigma_{t}^2\quad&else\\
		\end{aligned}
	\right.,
\end{equation}
to ensure that $\max\{ \sigma_{dp}^2,\sigma_{model}^2,\sigma_{channel}^2\} \geq \sigma_{t}^2$.

In brief, the proposed mechanism first confirms whether the channel noise and model noise are sufficient to achieve the differential privacy objective, and if not, it then chooses (\ref{eq_with_h}) or (\ref{eq_without_h}) to introduce the differential privacy noise in $\boldsymbol{x}$ based on whether or not it has $\boldsymbol{h}$.
Devices with better estimation capabilities on $\boldsymbol{h}$ can more accurately add noise.
Since the mechanism adds $\boldsymbol{n}_{dp}$ to signal symbols and symbols have natural upper and lower bounds, the sensitivity is easy to estimate. This simplifies the implementation of differential privacy and makes the mechanism broadly adaptable to different tasks without the need for task-by-task sensitivity analysis.

\section{Performance Evaluation}

To evaluate the performance of the proposed system, we implement it following the classical work DeepSC\cite{xie2021deep} for text transmission tasks.
The entire network parameter settings are summarized in Table \ref{tab_parameter}.
The semantic codec and channel codec of the system have the similar settings as DeepSC.
There are four Transformer encoder layers in the semantic encoder, and four Transformer decoder layers in the semantic decoder.
The channel codec consists of multiple layers of Dense.
The semantic knowledge network is the newly designed part in comparison to DeepSC.
The output of the semantic knowledge network is reshaped to $\mathbb{R}^{8\times128}$, as a semantic knowledge base.
The dataset used in evaluations is the English and French corpora in the proceeding of the European Parliament\cite{papineni2002bleu}.
The adopted datasets include English corpus, French corpus, and English-French corpus.
We use the vocabulary of the English-French corpus in word embedding for these three datasets.

We simulate the update phase of the proposed system.
The semantic knowledge network is froze and other parts of codecs are trained with the English-French corpus in the first 600 rounds, obtaining the initial codecs which supports both English and French text transmissions.
During the subsequent training, we train the initial codecs and the semantic knowledge network with different datasets, which represents the continuous update for different task requirements in the distributed system.
At this phase, the semantic and channel codecs are trained only $5$ rounds per $100$ rounds on average, while the semantic knowledge network is trained every round to generate the compressive semantic knowledge base.
On average, this training strategy greatly reduces the number of parameters being updated per round of the training, and therefore reduces the amount of data that needs to be shared during system update phase.
In all the above training processes, three channels with a SNR of $6dB$, AWGN channel, Rayleigh channel and Rician channel with $r=1$, are randomly selected.
The batch size is $64$, and Adam optimizer is adopted with an initial learning rate of $0.0001$, $\beta_1=0.9$ and $\beta_2=0.98$.
We set that $\{\lambda_1,\lambda_2\}=\{1,1\}$.
We measure the performance of the proposed system using the bilingual evaluation understudy (BLEU-1) score\cite{papineni2002bleu} by measuring the difference between words in two sentences.
\begin{table}[]
	\caption{The settings of the proposed system}
	\centering
	\renewcommand{\arraystretch}{1.5}
	\setlength{\tabcolsep}{5pt}
	\label{tab_parameter}
	\begin{tabular}{|c|c|c|}
		\hline
		& \textbf{Layer Name}          & \textbf{Unit}  \\ \hline
		\multirow{2}{*}{\textbf{Semantic Encoder}}       & Embedding                    & 128            \\ \cline{2-3} 
		& 4$\times$Transformer Encoder & 128 (8 heads)  \\ \hline
		\multirow{2}{*}{\textbf{Channel Encoder}}        & Dense                        & 256            \\ \cline{2-3} 
		& Dense                        & 16           \\ \hline
		\multirow{3}{*}{\textbf{Channel Decoder}}        & Dense                        & 128            \\ \cline{2-3} 
		& Dense                        & 512            \\ \cline{2-3} 
		& Dense                        & 128            \\ \hline
		\textbf{Semantic Decoder}                        & 4$\times$Transformer Decoder & 128 (8 heads)  \\ \hline
		\textbf{Predictable Layer}                       & Dense                        & Dictonary size \\ \hline
		\multirow{2}{*}{\textbf{Semantic Knowledge Net}} & Dense                        & 128            \\ \cline{2-3} 
		& Dense                        & 128$\times$8   \\ \hline
	\end{tabular}
\end{table}

\subsection{Performance with Compressive Semantic Knowledge Base}\label{evaluation_perform_CSKB}
\begin{figure}[t]
	\centering
	\includegraphics[width=3.5in]{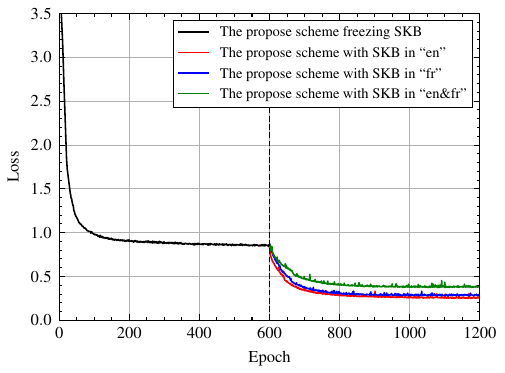}
	\caption{The evolution of test losses with semantic knowledge bases over AWGN wireless channels in a SNR of $12dB$.}
	\label{fig_p3}
\end{figure}

To demonstrate that the proposed update method with compressed semantic knowledge base achieves text transmission accuracy improvement during model update training in the dynamic environment, the evolution of test losses is shown in Fig. \ref{fig_p3}, where the loss is $\mathcal{L}_{CE}$ in (\ref{eq_loss_ce}).
``freezing SKB'' represents the process of training the initial codecs.
``SKB in `en' '', ``SKB in `fr' ''and ``SKB in `en\&fr' '' denote model update training with English corpus, French corpus, and English-French corpus respectively.
From Fig. \ref{fig_p3}, we know that at the $600$-th epoch, the loss has converged.
However, after the $600$-th epoch, the model update training using the semantic knowledge network achieves the lower loss convergence. 

\begin{figure*}[h]
	\centering
	\subcaptionbox{AWGN\label{fig_p1_AWGN_en}}
	{\includegraphics[width=2.3in]{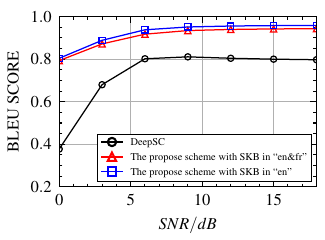}}
	\subcaptionbox{Rayleigh\label{fig_p1_Rayleigh_en}}
	{\includegraphics[width=2.3in]{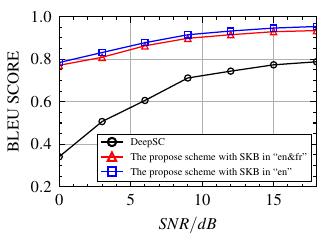}}
	\subcaptionbox{Rician\label{fig_p1_Rician_en}}
	{\includegraphics[width=2.3in]{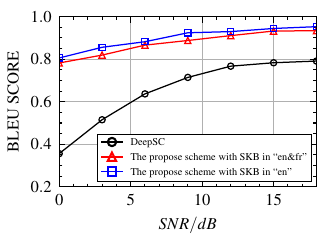}}
	\caption{Comparison of BLEU versus SNR for different $\boldsymbol{\kappa}$ with $100\%\boldsymbol{f}$ in English transmission task over different wireless channels.}	
	\label{fig_skb_e_3channel}
\end{figure*} 
\begin{figure*}[h]
	\centering
	\subcaptionbox{AWGN\label{fig_p1_AWGN_fr}}
	{\includegraphics[width=2.3in]{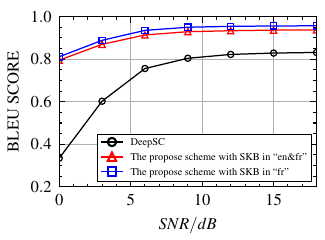}}
	\subcaptionbox{Rayleigh \label{fig_p1_Rayleigh_fr}}
	{\includegraphics[width=2.3in]{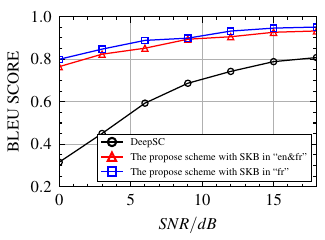}}
	\subcaptionbox{Rician\label{fig_p1_Rician_fr}}
	{\includegraphics[width=2.3in]{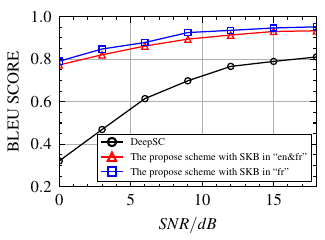}}
	\caption{Comparison of BLEU versus SNR for different $\boldsymbol{\kappa}$ with $100\%\boldsymbol{f}$ in French transmission task over different wireless channels.}	
	\label{fig_skb_f_3channel}
\end{figure*} 
To further show the effectiveness of the compressive knowledge base, we compare the BLEU versus SNR in English and French transmission tasks with different knowledge bases over various wireless channels, shown in Fig. \ref{fig_skb_e_3channel} and Fig. \ref{fig_skb_f_3channel}.
The DeepSC trained with English corpus and French corpus serve as baselines for comparisons.
Compared to the proposed system, the DeepSC is only lack of the semantic knowledge network.
For the proposed scheme, $100\%$ of semantic features are transmitted.
From the figures, it is observed that BLEU scores of the proposed scheme are higher compared to DeepSC in all the cases, especially when the SNR is low. Specifically, with a SNR of $0dB$, the proposed scheme still achieves BLEU scores of around $0.8$ in each of the three types of channels, whereas that of DeepSC is only around $0.3$. The reason why the proposed scheme achieves a significant advantage with a lower SNR is that the compressive knowledge base has been synchronized in advance and will not be affected by poor channel conditions during the communication process.
In addition, by comparing performances of different knowledge bases in the same task, it can be found that the closer the adopted training dataset is to the data to be transmitted in the task, the more the trained semantic knowledge base improves the BLEU.

We then evaluate the flexibility of the proposed semantic coding method.
Table. \ref{tab_trim} presents comparison of its BLEU under different pruning levels of $\boldsymbol{\kappa}$ and $\boldsymbol{f}$ with a SNR of $9dB$ over AWGN wireless channels. 
To reduce data exchanged, the knowledge base is pruned in this evaluation, leaving two parameters with the largest absolute values for each vector and setting the other parameters to zero.
The results show that as the level of pruning features increases, the BLEU of the system decreases, but the loss is compensated by utilizing the compressive semantic knowledge bases. 
Specifically, when using one knowledge vector and transmitting $40\%$ of semantic features, the BLEU score of the system is comparable to the DeepSC.
When the proposed system transmits the same amount of semantic features as DeepSC, the knowledge base improves the BLEU of the system by more than $16\%$.
In addition, we discover that a larger knowledge base is not always better.
With a small number of semantic features, the knowledge base becomes a major part of the semantic encoder input, which in turn reduces the BLEU score because the knowledge base holds the knowledge of the overall task rather than the information of a single sentence during a single transmission.

\begin{table*}[]
	\caption{Comparison of BLEU under different sizes of $\boldsymbol{\kappa}$ and $\boldsymbol{f}$ over AWGN wireless channels in a SNR of $9dB$, $\aleph$ ratio of improvement compared to BLEU of DeepSC.}	
	\centering
	\renewcommand{\arraystretch}{1.5}
	\label{tab_trim}
		
		\begin{tabular}{ccccccccc}
			\toprule
			& \renewcommand{\arraystretch}{1.0}\begin{tabular}[c]{@{}c@{}}BLEU Score \\ with $|\boldsymbol{\kappa}|=1$\end{tabular} & $\aleph$ &\renewcommand{\arraystretch}{1.0} \renewcommand{\arraystretch}{1.0}\begin{tabular}[c]{@{}c@{}}BLEU Score \\ with $|\boldsymbol{\kappa}|=2$\end{tabular} & $\aleph$ & \renewcommand{\arraystretch}{1.0}\begin{tabular}[c]{@{}c@{}}BLEU Score\\ with $|\boldsymbol{\kappa}|=4$\end{tabular} & $\aleph$ & \renewcommand{\arraystretch}{1.0}\begin{tabular}[c]{@{}c@{}}BLEU Score \\ with $|\boldsymbol{\kappa}|=8$\end{tabular} & $\aleph$ \\ \midrule
			$30\%\boldsymbol{f}$  & 0.7991                                                                                                                                                  & -1.49\%  & 0.7995                                                                                                                                                                                                                      & -1.44\%  & 0.7945                                                                                                                                                 & -2.06\%  & 0.7893                                                                                                                                                  & -2.69\%  \\
			$40\%\boldsymbol{f}$  & 0.8164                                                                                                                                                  & 0.65\%   & 0.8250                                                                                                                                                                                                                      & 1.70\%   & 0.8224                                                                                                                                                 & 1.38\%   & 0.8156                                                                                                                                                  & 0.55\%   \\
			$50\%\boldsymbol{f}$  & 0.8349                                                                                                                                                  & 2.93\%   & 0.8491                                                                                                                                                                                                                      & 4.67\%   & 0.8501                                                                                                                                                 & 4.80\%   & 0.8445                                                                                                                                                  & 4.10\%   \\
			$60\%\boldsymbol{f}$  & 0.8520                                                                                                                                                  & 5.03\%   & 0.8732                                                                                                                                                                                                                      & 7.65\%   & 0.8781                                                                                                                                                 & 8.25\%   & 0.8728                                                                                                                                                  & 7.59\%   \\
			$70\%\boldsymbol{f}$  & 0.8728                                                                                                                                                  & 7.60\%   & 0.8989                                                                                                                                                                                                                      & 10.82\%  & 0.9072                                                                                                                                                 & 11.84\%  & 0.9010                                                                                                                                                  & 11.08\%  \\
			$80\%\boldsymbol{f}$  & 0.8949                                                                                                                                                  & 10.32\%  & 0.9242                                                                                                                                                                                                                      & 13.94\%  & 0.9343                                                                                                                                                 & 15.18\%  & 0.9276                                                                                                                                                  & 14.35\%  \\
			$90\%\boldsymbol{f}$  & 0.9117                                                                                                                                                  & 12.39\%  & 0.9383                                                                                                                                                                                                                      & 15.67\%  & 0.9480                                                                                                                                                 & 16.87\%  & 0.9441                                                                                                                                                  & 16.39\%  \\
			$100\%\boldsymbol{f}$ & 0.9168                                                                                                                                                  & 13.02\%  & 0.9407                                                                                                                                                                                                                      & 15.97\%  & 0.9502                                                                                                                                                 & 17.14\%  & 0.9469                                                                                                                                                  & 16.73\%  \\ \bottomrule
			\end{tabular}
\end{table*}
\subsection{Update Efficiency with Compressive Semantic Knowledge Base}
To demonstrate that the proposed system greatly reduces the number of parameters updated in the model update training, without sacrificing the BLEU score and significantly increasing the complexity, we compare the proposed system with other methods in terms of updated parameters number per round, number of parameters, inference runtime per batch and BLEU score in the Table \ref{tab_main_usage}.
For fairness in the comparison, we select the Teacher model in \cite{liu2023knowledge} trained by English corpus with SNR varying randomly from $10 dB$ to $15 dB$ as the base model, on which other three methods are based for the model update training with SNR between $15 dB$ and $18 dB$.
The transfer learning (TL) method in \cite{xie2021deep} freezes the semantic codecs and only trains parts of the channel encoder and decoder.
Similar to the training strategy explained in \ref{evaluation_perform_CSKB}, our method trains the semantic knowledge network every round and other components are only updated once per 20 rounds.
The semantic knowledge net has only one layer of dense, where the input size is $8$ and the output size is $128\times8$.
In the knowledge distillation (KD) method\cite{liu2023knowledge}, a smaller model, Student 3, is trained with the help of the Teacher model through the knowledge distillation approach.
We evaluate BLEU scores in Rayleigh fading channels with a SNR of $18 dB$.
The comparison illustrates that our method achieves the highest BLEU scores with the least number of updated parameters, at the cost of a slight increase in overall model parameters and inference time.

\begin{table}[]
	\caption{The comparison of different methods in system update.}
	\label{tab_main_usage}
	\centering
	\renewcommand{\arraystretch}{2}
	\setlength{\tabcolsep}{1.7pt}
	\begin{tabular}{ccccc}
		\toprule
		                                                                                                                  & \renewcommand{\arraystretch}{1.0}\begin{tabular}[c]{@{}c@{}}Updated parameters\\ per round\end{tabular} & Parameters & Runtime & BLEU \\ \midrule
		Base Model                                                                                                             & -                                                                & 2022672    & 90ms   & -      \\
		TL in \cite{xie2021deep}                                                              & 171280                                                            & 2022672    & 90ms   & 0.904         \\
		KD in \cite{liu2023knowledge}                                                    & 946704                                                                 & 946704     & 55ms    & 0.907      \\
		\renewcommand{\arraystretch}{1.0}\begin{tabular}[c]{@{}c@{}}Ours with\\ $75\%\boldsymbol{f}$ and $|\boldsymbol{\kappa}|=8$\end{tabular} & 110350                                                                 & 2031888    & 91ms   & 0.945          \\ \bottomrule
	\end{tabular}
\end{table}
\subsection{Performance with the Signature Mechanism for Lossy Semantics}
\begin{figure}[t]
	\centering
	\includegraphics[width=3.5in]{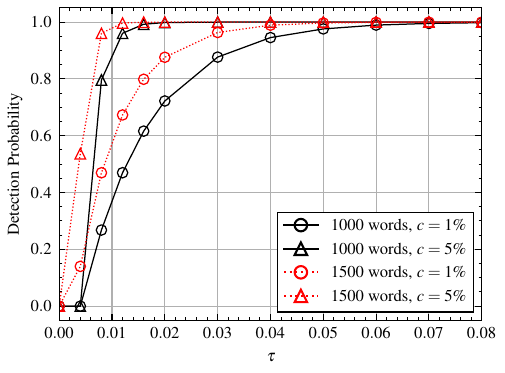}
	\caption{$\tau$ versus detection probability for different ratios $c$ of corrupted parameters in $\boldsymbol{d}$.}
	\label{fig_overhead}
\end{figure}

This section presents that the proposed signature mechanism achieves a high probability of detecting semantics tampering when $\eta_i>\bar\eta$ while introducing less additional communication burden.
In semantic communications with finite constellation, for a block $\boldsymbol{d}$ with $I/2$ symbols to be transmitted, the signature mechanism requires the extra transmission of $S=\left\lbrace {\boldsymbol{d}}_{\boldsymbol{\rho}}  ||n||s \right\rbrace$.
We select DSA with a key length of $1024$ bit to implement $sig$, so that the length of signature $s$ is $1024$ bit.
Considering the above parameters are 32-bit floating points, 
$S$ can be assumed to have $\left( |\boldsymbol{\rho}|+1+(1024/32)\right) $ floats.
The additional communication cost is defined as
\begin{equation}
	\tau \triangleq \frac{|\boldsymbol{\rho}|+1+(1024/32)}{I/2} float / symbol.
\end{equation}
Fig. \ref{fig_overhead} shows the additional communication cost $\tau$ versus detection probability for different ratios of corrupted parameters in $\boldsymbol{d}$, denoted as $c$.
The experiments are conducted in two settings where $\boldsymbol{d}$ are signal symbol parameters extracted from $1000$ or $1500$ words texts, which are lengths of common articles.
From the Table \ref{tab_parameter}, we derive that the channel encoder generates $16/2$ signal symbols for each word.
The results indicate that even if only $1\%$ of the semantics is tampered with, the proposed signature mechanism achieves more than $95\%$ probability of detecting semantics tampering or forgery while introducing no more than $5\%$ addition communication cost.
\subsection{Performance with the Noise-Aware Differential Privacy Mechanism}
\begin{figure}[t]
	\centering
	\includegraphics[width=3.5in]{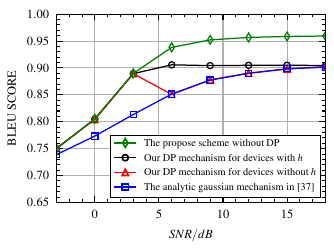}
	\caption{Comparison of BLEU versus SNR over AWGN channel for different DP mechanism with $\epsilon=2$ and $\delta=0.05$}	
	\label{fig_eval_dp}
\end{figure}

To show that proposed privacy-preserving mechanism achieves better communication performance by optimizing the differential privacy noise, we compare their scores of BLEU with the traditional approach at the same DP setting of $\epsilon=3$ and $\delta=0.05$ in Fig. \ref{fig_eval_dp}.
The traditional approach refers to adding Gaussian noise to the symbols directly based on $\epsilon$ and $\delta$ via the analytic gaussian mechanism \cite{balle2018improving}, without considering model noise and channel distortions.
Evaluations are performed with model noise unavailable.
The line with `$\triangle$' indicates the DP mechanism based on (\ref{eq_with_h}) when $\boldsymbol{h}$ is known, and the line with `$\square$' represents the implementation of the DP mechanism based on (\ref{eq_without_h}) when $\boldsymbol{h}$ is unknown.
From the results, it can be seen that BLEU scores of the two proposed mechanisms are always higher than that of the traditional DP mechanism.
When the $SNR$ is below $3 dB$,  their BLEU scores are optimal because the two mechanisms do not introduce extra noise and BLEU scores are the same as the proposed scheme without DP.
In addition, the mechanism with (\ref{eq_with_h}) achieves a higher BLEU score than the mechanism with (\ref{eq_without_h}) because it has additional channel information $\boldsymbol{h}$ to better optimize the differential privacy noise.
The results show that proposed DP mechanisms are able to guarantee mathematically rigorous proofs of privacy preservation with better communication performance compared to traditional approach.
This is due to two important reasons, firstly, the proposed differential privacy mechanism reduces the added noise required to achieve differential privacy, and secondly, the system uses a semantic knowledge base to compensate for the loss of performance due to the addition of noise.

\section{Conclusion}
We explore the security and practical deployment of the semantic communication system in distributed IoT networks in both update and usage phase.
A blockchain-based scheme for the trustworthy system update is designed, ensuring the integrity and availability of the update date shared between IoT devices.
The efficiency of system update is further improved by the proposed flexible and efficient semantic coding method base on compressive semantic knowledge base.
It achieves a better BLEU score compared to related works by updating only $5.43\%$ of all parameters per round on average in the model update training, which consequently reduces the amount of data exchange required for system update.
The method also achieves a flexible model training and inference for heterogeneous devices, supporting the adjustment of the size of transmitted symbols and the knowledge base.
In the usage phase, we develop a signature mechanism to verify the integrity and authenticity of lossy semantics. The effect of wireless channels on transmitted semantics are evaluated in the verification process.
It realizes high probability of detecting semantics tampering with a small additional transmission burden.
We further introduce a noise-aware differential privacy mechanism to defend against malicious data analysis.
The mechanism analyze the lossy transmission characteristics of semantic communications to optimize the additional noise required to achieve differential privacy.
The availability of channel information and model noise information is taken into account to provide diverse implementations for heterogeneous devices.
Therefore, the proposed system is an attractive potential solution for secure and efficient intelligent IoT networks.

\bibliographystyle{IEEEtran}
\bibliography{ref}

\end{document}